\documentstyle[sprocl]{article}

\def\be{\begin{equation}}
\def\ee{\end{equation}}
\def\bea{\begin{eqnarray}}
\def\eea{\end{eqnarray}}

\begin{document}

\title{REVIEW OF ATMOSPHERIC NEUTRINOS}

\author{ T.K. GAISSER}

\address{Bartol Research Institute, University of Delaware\\
Newark, DE 19716, USA}


\maketitle
\abstracts{
In this talk I review  measurements and calculations of
the flux of neutrinos produced by interactions of cosmic
rays in the atmosphere.  The main reason for interest in
this subject is the apparent anomaly between the predicted
and the observed ratio of $\nu_e$ to $\nu_\mu$.  With the
advent of Super-Kamiokande, we are
on the threshold of an order of magnitude increase in the
amount of data available to study the problem.  My goal
in this talk is to describe the current status of the
subject, both for contained events and for neutrino-induced
upward muons.}

\section{Introduction}
Because of their small cross sections
neutrinos were the last component of the secondary cosmic radiation
to be measured, although they are the most numerous particles
in the GeV energy range at sea level.  Markov~\cite{Markov}
suggested how upward and horizontal muons deep underground could be
used as a signal of high energy neutrinos, and Greisen~\cite{Greisen}
described a neutrino detector like the modern water detectors
in which neutrino interactions could be observed directly.\cite{RoySoc}
Neutrino-induced horizontal muons at the predicted level
were observed a few years later in deep mines in India~\cite{Achar}
and in South Africa.\cite{Reines}

The deep detectors built to search for
proton decay have now accumulated more than a thousand
contained events.  The
large water Cherenkov detectors, IMB~\cite{IMB} and Kamiokande,\cite{Kam}
dominate the statistics.  Although the total
events rates are consistent with the expectation for
interactions of atmospheric neutrinos, the ratio of electron-type
to muon-type neutrinos is significantly higher than predicted
for the water Cherenkov detectors.\cite{IMB,Kam}  The first
hint of this anomaly came already in 1986 with the observation
by IMB of fewer than expected muon decays among their
events.\cite{Haines}   The most well-defined class of events is
the contained single-ring events.  These are mostly
charged-current quasi-elastic events (e.g. $\nu_\mu + n\rightarrow
p + \mu^-$) with with an admixture of neutral-current events
in which a single pion is produced.  A recent statement of
the anomaly for $\le GeV$ neutrinos at Kamiokande is~\cite{Suzuki}
$$
{(\mu/e)_{\rm data}\over (\mu/e)_{\rm calculated}}\;=\;0.60_{-.05}^{+.06}
$$
for contained single-ring events.
The iron tracking calorimeters, however, find
results consistent with no anomaly~\cite{Frejus,NUSEX} or with
a smaller anomaly.\cite{Soudan,Peterson}

The rate of interactions for neutrinos $\nu_i$ inside a detector of
mass M (in g) is
\begin{equation}
Rate =N_A\,M\, \int dE_\nu\int dE_\ell\int d\Omega\,
\phi_i(E_\nu,\Omega)\,{d\sigma_i\over dE_\ell}\,\epsilon(E_\ell),
\label{Rate}
\end{equation}
where $N_A$ is Avogadro's number. The factors in the
integrand are the differential flux of $\nu_i$; the differential
cross section to produce the corresponding lepton, $\ell$; and the
efficiency, $\epsilon$, for its identification and detection.
The expression for the rate of neutrino-induced muons is similar
except that the target mass is \begin{equation}
M(\Omega)\;=\;R_\mu(E_\mu)\times A(\Omega),
\end{equation}
where $R_\mu(E_\mu)$ is the muon range and $A(\Omega)$ is the
projected area of the detector as seen from the direction
$\Omega = (\theta,\phi)$.

Explanations for the anomaly in the
contained events have been sought in all three factors of Eq.~\ref{Rate}.
I will discuss calculation of the flux $\phi_i$ of atmospheric
neutrinos separately in the next section.  
Several approximations have been made in the treatment of
the cross sections for neutrino interactions in oxygen
inside the water detectors.  Engel et
al.\cite{Vogel} conclude, however, that the neglected physics
cannot account for the anomalous $\mu$-to-$e$ ratio observed at
Kamiokande and IMB.  The reason is that the lepton momenta are
high enough that corrections affect muons and electrons in very
nearly the same way.
Beam tests at KEK~\cite{beamtest,Totsuka} confirm the efficiencies determined
from simulations for misidentifying muons as electrons and vice versa.

Another possibility is that there is some contamination of
events that are not due to interactions of atmospheric neutrinos.
The suggestion that there was an excess of events due
to $p\rightarrow \nu\bar{\nu}e$~\cite{TonyMann} is presumably
eliminated by the fact that the anomaly persists at higher energy.\cite{Fukuda}
Ryazhskaya~\cite{Ryazhskaya} has suggested that extra electron-like
events are really cascades from neutral pions produced inside
the detector by interactions of entering neutrons generated
outside the detector in interactions of atmospheric muons in the rock.
Kamiokande argues against this by showing~\cite{Totsuka} that they have a
$\pi^0$ production rate consistent with neutral current
production by neutrinos at the level expected and already
accounted for in their simulations.  Soudan
argues against this explanation by analysis of their shield
events.\cite{Soudan,Peterson}  

Such possibilities highlight the importance of internal checks
on the data.  Various analyses~\cite{Beieretal,GHS} agree
that the Kamiokande and IMB data are consistent with each other
after accounting for differences in exposure, geomagnetic cutoff and
energy threshold, although Beier and Frank~\cite{Beier} note a
possible small discrepancy in the electron spectra at low momentum.  
There is a hint of the expected smaller interaction
rate at the IMB detector during the period of maximum solar
activity between 1989 and 1991 as compared to the earlier
part of their data collection from 1986-88, a period of minimum
solar activity. One expects fewer events during solar maximum
when the low energy primary cosmic rays are partially excluded from the
inner solar system.  The effect should be more noticeable
in the overall rate at IMB than at Kamiokande because of
the higher local geomagnetic cutoff at Kamiokande, which excludes
a large fraction of the lower energy primaries in any case.
During the first period (exposure of 3.4 kT yrs)
IMB found 236 events (139 e-like and 97 $\mu$-like).\cite{Casper}
The corresponding numbers for the last 4.3 kT yrs of the 7.7 kT-yr.
exposure were~\cite{IMB} 271 (186 e-like and 85 $\mu$-like).
The total expected by simple extrapolation of the first
period would be $297\pm 20$.  The difference ($297-271$),
though not statistically significant, is about what would 
be expected due to solar modulation effects.  The fact that
the decrease shows up only in the $\mu$-like events
is not consistent with solar modulation.  Presumably it
is an accident of low statistics.
Stanev~\cite{Stanev} has emphasized
the importance of using the expected solar cycle variations
as a probe of atmospheric neutrino data, which will be
possible with larger data samples.

\newpage

Before turning to the discussion of the neutrino fluxes, I
display the integrand of Eq.~\ref{Rate}.  Fig. 1 shows
the ``response-curves'' for observed neutrino-induced muons.
What is plotted is the rate of muons integrated over muon
energy per logarithmic interval of neutrino energy.
The four classes of events are fully-contained interactions,
Kamiokande ``multi-GeV'' events,\cite{Fukuda} entering muons
that stop in the detector and neutrino-induced throughgoing muons.
The figure~\cite{FGMS} is specific to the Kamiokande detector, but
similar distributions could be constructed for any detector.  The curves
rise from low energy as the neutrino cross section
increases with energy.  For $\nu$-induced muons the effective
volume also increases with energy as the range increases.
Eventually the growth of range and cross section slow and
the steep cosmic-ray spectrum (together with pion interaction)
cuts off the signal of atmospheric neutrinos at high energy.
The response curves are useful when considering possible
explanations of the flavor anomaly in terms of neutrino oscillations.

\section{Flux of atmospheric neutrinos}

The analysis of atmospheric neutrino experiments has depended
mainly on four calculations of atmospheric neutrinos.\cite{BGS,HKHM,LK,BN}
The calculations of G. Barr, Gaisser and Stanev (BGS)~\cite{BGS},
Honda et al. (HKHM)~\cite{HKHM} and Bugaev and Naumov (BN)~\cite{BN}
are completely independent of each other.
The neutrino flavor ratio is the same within 5\% in all
these calculations, but there are some significant differences
in normalization and shape of the calculated neutrino
energy spectra.  Many sources of uncertainty cancel in
the calculation of the ratio of $\nu_e/\nu_\mu$.
Thus the theoretical uncertainties are much smaller in the
ratio than in the normalization.  Fogli and Lisi~\cite{FL}
have shown how to make the comparison between expected
and measured neutrino interactions in this situation.

Suzuki~\cite{Suzuki} has compared the measured spectra of
electrons and muons in single ring neutrino interactions
with full simulations starting from the fluxes of
HKHM, BGS and BN.
The calculation of BGS~\cite{BGS} gives the steepest
spectrum, predicting more muons with momenta below
600 MeV/c than observed, but agreeing with the measured spectrum of
electrons.  In contrast, the neutrino spectra of BN~\cite{BN}
are nearly in agreement with the muon spectrum down to
200 MeV/c but predict fewer electrons than observed.  The calculation
of HKHM~\cite{HKHM} is intermediate but closer to the
results of BGS~\cite{BGS}.

Table 1~\cite{GHKLMNS} compares
the neutrino fluxes of the three calculations.  The first
part of the table shows the neutrino spectra separately for
$\nu_e$ and $\nu_\mu$ in three energy
intervals, normalized to BGS$=1.00$.  The second part
of the table shows the
neutrino ratios in the energy interval between 0.4 and 1 GeV.    Here
$$R_{e/\mu}\;\equiv\;
{\nu_e+{1\over 3}\bar{\nu}_e\over\nu_\mu+{1\over 3}\bar{\nu}_\mu}
$$
to reflect the smaller interaction cross section for antineutrinos.
This crucial ratio is nearly the same in all cases.

\begin{table}
\caption{Comparison of calculated neutrinos fluxes at Kamioka}
\begin{center}
\vspace*{1truemm}
\begin{tabular}{|l|ccc|ccc|}  \hline
     & $\nu_\mu+\bar{\nu}_\mu$ &&& $\nu_e+\bar{\nu}_e$ && \\
     & 0.4 -- 1 & 1 -- 2 & 2 -- 3 & 0.4 -- 1 & 1 -- 2 & 2 -- 3  \\ \hline
BGS  & 1.00 & 1.00 & 1.00 & 1.00 & 1.00 & 1.00 \\
HKHM & 0.90 & 0.95 & 1.04 & 0.87 & 0.91 & 0.97 \\
 BN  & 0.63 & 0.79 & 0.95 & 0.62 & 0.74 & 0.87 \\ \hline
\end{tabular}
\vspace*{5truemm}
\begin{tabular}{|l|ccc|}                                        \hline
  &      $\bar{\nu}_\mu/\nu_\mu$  &  $\bar{\nu}_e/\nu_e$ & $R_{e/\mu}$ \\
        & $0.4$ & $\le\,E_\nu$ & $\le\,1$~GeV                     \\ \hline
BGS & 0.99 & 0.89 & 0.49  \\
HKHM  & 0.99 & 0.84 & 0.48  \\
 BN  &  0.98 & 0.76 & 0.50  \\ \hline
\end{tabular}
\end{center}
\end{table}

We~\cite{GHKLMNS} have investigated the sources of difference
among the calculations by substituting one-by-one different
sets of assumptions from the various papers into the framework
of the BGS calculation.   In that work~\cite{BGS}
the spectrum of neutrinos
$\nu_i$ was expressed as a convolution of the primary
cosmic-ray spectrum, the geomagnetic cutoffs and the yield
per nucleon of $\nu_i$:
\begin{eqnarray}
\label{nuflux}
\phi_{\nu_i}(\Omega)\;=&\phi_p\otimes R_p\otimes
Y_{p\rightarrow\nu_i}\\ \nonumber
  &+\;\phi_{p(A)}\otimes R_A\otimes Y_{p\rightarrow\nu_i}\\ \nonumber
  &+\;\phi_{n(A)}\otimes R_A\otimes Y_{n\rightarrow\nu_i}.
\end{eqnarray}
In this equation $R_p(R_A)$ is the geomagnetic cutoff for
protons (nuclei) incident on the atmosphere from the direction
$\Omega = \{\theta,\phi\}$.  The three terms on the right-hand-side
represent respectively neutrinos from primary hydrogen, from
protons bound in incident nuclei and from neutrons bound in incident
nuclei.  The separation is necessary because neutrino production
depends on energy-per-nucleon of the incident cosmic rays but
the geomagnetic cutoff depends on magnetic rigidity (gyroradius).
Nuclei and protons of the same energy per nucleon
differ by a factor of $A/Z$ in magnetic rigidity.
To a good approximation~\cite{Engel} the yields of neutrinos
from nuclei can be calculated as if the incident nucleons were unbound.
(This approximation somewhat overestimates the production
of neutrinos from pions produced in the target fragmentation
region for the fraction of the flux due to incident nuclei.
It is not used in Ref.~\cite{BN}.)
In the energy region important for contained events, approximately
80\% of the neutrinos are produced by cosmic-ray hydrogen (free
protons) and most of the rest come from helium nuclei.

The form of the BGS calculation (Eq. \ref{nuflux}) makes it possible
to trace the effects of different assumptions through the
calculation.  As an example, consider the 1 GeV neutrino
flux at Kamiokande at solar minimum.  (Kamioka has the
highest cutoff for downward cosmic rays of the nucleon decay
detectors, so the effect of differences in cutoff is maximum.
Comparison at solar minimum maximizes the effects of differences
in assumed primary spectrum.)  We compare the calculation
of BGS~\cite{BGS} with that of HKHM~\cite{HKHM}.  The treatment
of the geomagnetic cutoff in BGS neglected the
``penumbra'' effect.  A more accurate treatment of the cutoffs
in by HKHM reduces the 1 GeV flux by a factor 0.87.
The primary spectrum assumed in HKHM is higher
than in BGS, which gives a factor 1.25.  Yields
of GeV neutrinos are approximately 15\% lower in HKHM,
giving a factor of 0.85.  The product $0.87\times1.25\times0.85=0.93$
gives a net 7\% lower GeV neutrino flux in HKHM
than in BGS.

The main result of our comparison~\cite{GHKLMNS} is that
the biggest source of difference among the three independent
calculations~\cite{BGS,HKHM,BN} is the treatment of production
of low energy pions in collisions of 10 to 30 GeV protons with
nuclei of the atmosphere.  In the
BN calculation the inclusive cross sections
for production of $<3$~GeV pions is significantly lower than
in the calculations of BGS and HKHM.  In fact, it is quite
similar to the spectrum of pions in proton-proton interactions.
In contrast, both HKHM and BGS use representations of
pion production that give significantly more low energy pions
in collisions on nitrogen than for $pp$ collisions.
It is this feature of the BN calculation that gives rise
to their characteristically harder neutrino spectra with
relatively few low energy
($<2$~GeV) neutrinos.  At higher energy the neutrino fluxes of
the three calculations are in better agreement.

Measurements of muons at high altitude can be used to
constrain the neutrino spectra.  Perkins~\cite{Perkins} has calculated
neutrino spectra starting from measured muon spectra,
including preliminary results of the MASS~\cite{Circella} experiment.
He concludes that the higher flux is preferred.
Several groups have now measured
the muon flux during the ascent of their balloon-borne detectors.
The spectra of negative muons first reported in Ref.~\cite{Circella}
have now been published.\cite{Bellotti}
The IMAX experiment has given a preliminary report of
their measurements of muons~\cite{Kriz}, and the HEAT experiment
is in the process of analyzing their measurements.\cite{Barwick}

Both the Japanese group~\cite{Honda} and we~\cite{Agrawal,GS}
have now published calculations of the neutrino fluxes
over the whole energy
range from $<100$~MeV to $10^4$~GeV.  Our calculations~\cite{Agrawal,GS} now
include a correct treatment of the geomagnetic cutoff effects,
as described by Lipari and Stanev.\cite{LS}
Because of the large range of energies they cover, these
calculations~\cite{Honda,Agrawal,GS} each give a consistent
set of neutrino fluxes for simulating the full range of
experimental data, from contained events to neutrino-induced muons
at high energy.  The algorithms used for these and other calculations of the
atmospheric neutrino flux should be checked by comparing
their corresponding muon fluxes with the full set
of measurements of muons at high altitude that are becoming available.

\section{Neutrino oscillation interpretation of the flavor anomaly?}

As shown in the previous section, the predicted neutrino flavor ratio
is quite robust because many sources of uncertainty cancel in
the ratio.    The most intensively
investigated physics explanation of the atmospheric neutrino anomaly
 is the possibility of neutrino
oscillations.  As an example, consider $\nu_\mu\leftrightarrow\nu_\tau$
oscillations, which occur with probability
\begin{equation}
P_{\nu_\mu\rightarrow\nu_\tau}\;=\;\sin^22\theta\,\sin^2\left(1.27\,
\delta m^2{L(\rm km)\over E(\rm GeV)}\right).
\label{nuosc}
\end{equation}
There is no visible up/down difference in contained events,
for which $E_\nu\sim 1$~GeV.  Since $L\sim 20$~km for downward
neutrinos this gives a lower limit of approximately,
$\delta m^2 \ge 0.005$~eV$^2$.
If the effect begins to disappear for higher energy, there would
then be an upper limit on $\delta m^2$.  This is the significance
of the Kamiokande extended analysis that includes a multi-GeV
sample of data.   In the multi-GeV sample there is an apparent
up/down difference in the comparison between calculated and
observed flavor ratio.  The anomaly is largest for upward neutrinos
with pathlength $L\sim R_\oplus\approx 6000$~km.  The ratio is consistent
with expectation for vertically downward events.

The Kamiokande combined analysis defines allowed regions
for both $\nu_e\leftrightarrow\nu_\mu$ and $\nu_\mu\leftrightarrow\nu_\tau$
oscillations with large mixing angle and $\delta m^2$
in the range 0.01 to 0.02 eV$^2$.
The preferred interpretation depends both on the normalization
and the shape of the calculated neutrino flux.
For the contained~\cite{IMB,Kam} and sub-GeV~\cite{Fukuda}
events the calculations with high normalization~\cite{BGS,HKHM}
suggest oscillations primarily in the $\nu_\mu\leftrightarrow\nu_\tau$
sector whereas the calculation of BN~\cite{BN}
prefers $\nu_\mu\leftrightarrow\nu_e$.

There is an interesting energy-dependence that shows up
in the multi-GeV events,\cite{Fukuda} which has been noted previously
(e.g. in Refs.~\cite{GG} and~\cite{FL}).  When compared to
the HKHM neutrino fluxes, the ratio of measured/calculated
for electron-like events is $\sim1.5$ as compared to $\sim1.1$
for the sub-GeV sample.  The corresponding numbers for muon-like
events are $\sim 0.83$ and $\sim 0.66$.  A neutrino spectrum
that is sufficiently harder could be made to give the same
ratios of measured/calculated for the sub-GeV and multi-GeV samples.\cite{FL}
(The statistical significance of the anomaly would not be changed,
only its interpretation.)  Inspection of Table 1 shows that the
BN spectrum has roughly the right degree of hardness to keep the ratio
of measured/calculated constant for each neutrino flavor, 
but at the cost of a rather
extreme assumption about the nature of production of low-energy
pions in collisions of 10---30 GeV protons on oxygen; namely, that
the yield is about the same below $\sim2$~GeV as for collisions
on protons.
In a recent comprehensive three-flavor treatment of the
atmospheric neutrino anomaly, Fogli et al.\cite{Foglietal}
find that in fact $\nu_\mu\leftrightarrow\nu_e$ oscillations
provide a better fit to the atmospheric neutrino
anomaly than $\nu_\mu\leftrightarrow\nu_\tau$.

\section{Neutrino-induced upward muons}

Another way to explore higher neutrino energies is to
use upward muons.  Table 2 summarizes some comparisons between
experiments and calculations for the flux of
neutrino-induced muons with energy greater than a few GeV
(the exact value depends on the experimental cut).  The
calculations are shown for two different assumed neutrino
spectra, Volkova~\cite{Volkova} and Bartol,\cite{Agrawal}
and assuming no oscillations.
The calculations also depend on the structure functions used
to obtain the cross section for $\nu\,+\,N\rightarrow\,\mu\,+ X$.

\begin{table}
\caption{Status of $\nu$-induced upward muons}
\begin{center}
\vspace*{1truemm}
\begin{tabular}{l|llll}
     & IMB \cite{IMB2} & Baksan \cite{Baksan}& Kamioka$^{(a)}$
     \cite{Suzuki} & MACRO \cite{Ronga} \\ \hline\hline
Observed & 617 & 559 & 1.97$\pm0.10$ & 255 \\ \hline
Calculated & & & & \\
 \, Bartol flux \cite{Agrawal} & --- & 580 & 2.15 & 315 \\
 \, Volkova flux \cite{Volkova} & 600 & --- & 2.06 & 286 \\ \hline
\end{tabular}
\end{center}
$^{(a)}$Entries for Kamiokande give fluxes in units of
$10^{-13}$cm$^{-2}$s$^{-1}$sr$^{-1}$.  Other experiments
quote total number of events.
\end{table}

The IMB calculation is an early result~\cite{IMB2} which uses the
EHLQ~\cite{EHLQ} structure functions which probably give too low
a value of the cross section and hence underestimate the expected
rate by some amount~\cite{FGMS}.  Baksan~\cite{Baksan} and MACRO~\cite{Ronga} 
both calculate the cross section using Morfin \& Tung
structure function B1-DIS~\cite{MT}, while Kamiokande~\cite{Suzuki}
use MRS set G.\cite{MRS}  The results are inconclusive
because interpretation depends on comparison with an absolute calculaton.
For example, Frati et al.\cite{FGMS} considered an oscillation purely
in the $\nu_\mu\leftrightarrow\nu_\tau$ sector.  Using 
Owens~\cite{Owens} structure function for the cross section, they found
1.61, 1.97, and $2.33\times 10^{-13}$cm$^{-2}$sr$^{-1}$s$^{-1}$
for $\delta m^2 = 0.01$~eV$^2$ and $\sin^2(2\theta)=1.0$, $0.5$ 
and $0.0$ (no oscillations) respectively.

  Comparison to the muon
flux is also relevant for high energy neutrino fluxes, though
the constraint becomes less restrictive as energy increases
because at high energy ($>100$~GeV) a relatively larger
fraction of neutrinos comes from kaon-decay as compared to
the muons, which are always dominated by pion decay.\cite{Lipari,GLS}
The muon fluxes corresponding to several different
calculations of the neutrino flux at high
energy~\cite{Volkova,Mitsui,Butkevich} are
compared with a compilation of measured vertical muon
fluxes from $1$ to $10^4$~GeV in Ref.~\cite{RoySoc}.
All these calculations show some tendency to be higher than
the mesaurements from $10$ to $100$~GeV  and somewhat
below the data from $100$ to $1000$~GeV.
This small systematic effect, which is also noted in Ref.~\cite{Agrawal}
is not presently understood.

The angular dependence of the upward (neutrino-induced)
muon flux also contains
in principle information relevant to an oscillation interpretation
because the pathlength changes with angle.  There are, however,
significant systematic uncertainties because the experiments
in general have acceptances that depend on direction.
(See for example the discussion of the angular-dependence
of the MACRO data.\cite{Ronga})
There is also some difference in angular dependence among the calculations.
In this situation it might be useful to carry out a two-dimensional
analysis of the comparison between data and calculation,
following the example of Fogli and Lisi~\cite{FL}
for contained events.  The
variables might be total rate and
the ratio of ``horizontal''/``upward.''

Concerning angular dependence, it is interesting to note
that the recent Kamiokande data set (364 events)~\cite{Suzuki} does not fit
any calculation as well as the earlier set~\cite{KAM2} with
poorer statistics (252 events).
For example, the $\nu_\mu\leftrightarrow\nu_\tau$
oscillation referred to above with $\delta m^2 = 0.01$ gives
reduced $\chi^2$ values of 2.9, 2.6, and 3.2 respectively
for $\sin^22\theta = 1.0,\,0.5$, and $0$ (no oscillation).
The corresponding $\chi^2$ values for the earlier data set
were 1.9, 1.1 and 2.0 (no oscillation).

Another way to remove some of the model-dependence from
a comparison between expectation and observation of neutrino-induced
muons is to compare the ratio of stopping to throughgoing muons.\cite{IMB2}
The stopping muons depend significantly on the neutrino cross
section in the few GeV region, which is below the deep-inelastic
scattering regime.  Lipari et al.\cite{LLS} have pointed out
that the cross section is poorly known in the low energy region.
They make a careful evaluation of the cross section in the
resonance region and find that the exclusion region is less
restrictive than originally estimated in Ref.~\cite{IMB2}.
There is almost no overlap between the excluded region they
find and the ``allowed'' region for oscillations in the
$\nu_\mu\leftrightarrow\nu_\tau$  sector found by Kamiokande.\cite{Fukuda}

\section{Conclusion}

Results from the two large water detectors~\cite{IMB,Kam} are consistent with
each other, and they report a significant anomaly in the ratio
of e-like to $\mu$-like events as compared to what is expected
if all the events are due to interactions of atmospheric neutrinos
inside the detectors.  Iron (tracking) detectors~\cite{Frejus,NUSEX}
are consistent with no anomaly but with low statistics. Preliminary
results of the Soudan experiment are intermediate.\cite{Soudan,Peterson}  
Calculated
ratios of $\nu_e/\nu_\mu$ are the same within $\pm5$\% in all
calculations.  Main sources of difference in overall normalization
and shape among the calculations
of the flux of atmospheric neutrinos have been identified.\cite{GHKLMNS}
There are updated calculations~\cite{Honda,Agrawal,GS} which cover the
whole neutrino energy range from $<100$~MeV to $10^4$~GeV.  Further
work is needed to compare these calculations with 
measurements~\cite{Bellotti,Kriz,Barwick}
of muons in the atmosphere at various cutoffs, times and altitudes. 

The angular dependence of the Kamiokande multi-GeV events,\cite{Fukuda} in
combination with their sub-GeV events,\cite{Kam} suggests 
a limited range of oscillation parameters with large mixing angles
and $\delta m^2$ in the range $\sim 10^{-2}$~eV$^2$.
Because they require comparison to an absolute calculation
(rather than a ratio as for the contained events), the  measurements
of neutrino-induced upward muons~\cite{Baksan,Ronga,KAM2,Suzuki}
are inconclusive at present.  

First results from Super-Kamiokande are expected very soon.\cite{YSuzuki}
Because of its large fiducial volume Super-K will be able to
accumulate of order $10^4$ events in a few years.  This is enough to give
$\sim 100$ events from each cone of half-angle $10^\circ$ in
the sky.  Super-K should therefore be able to pick out directions
with particularly high and particularly low values of geomagnetic
cutoff and therefore demonstrate that they can see the
appropriate changes of rate characteristic of 
neutrinos produced by cosmic rays in the atmosphere.  
Similar remarks can be made
about effects of solar modulation as we go from the current epoch of
minimum solar activity into the next solar maximum in $\sim 1999$.
In addition the systematics will improve as higher energy events
will be fully contained.

\section*{Acknowledgments}
I am grateful to Paolo Lipari, G.L. Fogli, Todor Stanev, Francesco Ronga
and S. Mikheyev for discussions and data that helped me prepare this talk.
This work is supported in part by the U.S. Department of Energy
under Grant No. DE-FG02-91ER40626.

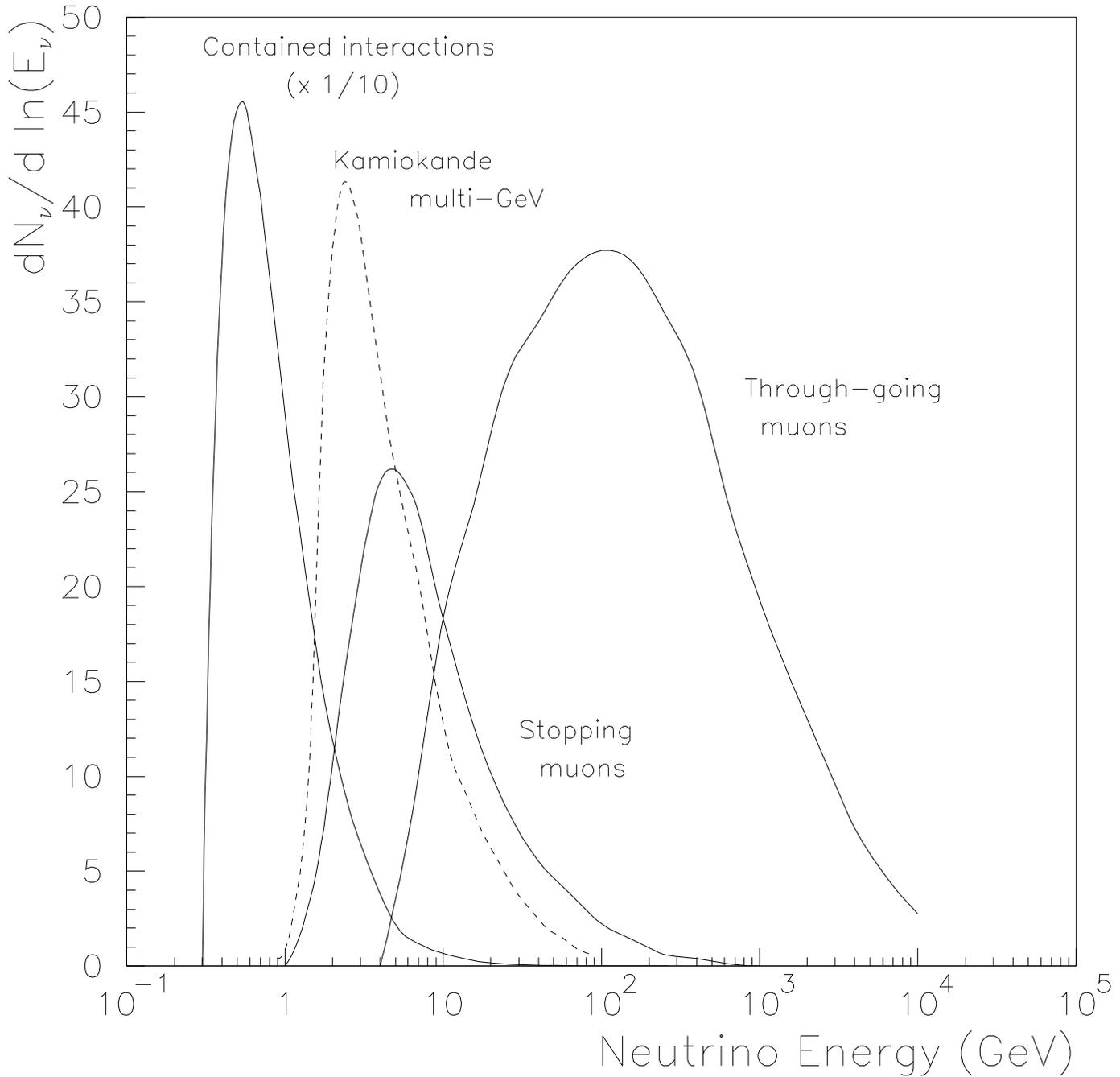
\end{document}